\documentclass[fleqn,usenatbib,13pt]{mnras}

\usepackage{newtxtext,newtxmath}
\usepackage{verbatim}
\usepackage{color}

\usepackage[T1]{fontenc}

\DeclareRobustCommand{\VAN}[3]{#2}
\let\VANthebibliography\thebibliography
\def\thebibliography{\DeclareRobustCommand{\VAN}[3]{##3}\VANthebibliography}

\usepackage{graphicx}	
\usepackage{amsmath}	

\title[The spin of 4U 1543-47]{The spin of a stellar black hole 4U 1543-47 determined by \textit{Insight}-HXMT}

\author[Chen \& Wang]{
Jiashi Chen,$^{1,2}$
Wei Wang$^{1,2}$\thanks{E-mail: wangwei2017@whu.edu.cn}
\\
$^{1}$Department of Astronomy, School of Physics and Technology, Wuhan University, Wuhan 430072, China \\
$^{2}$WHU-NAOC Joint Center for Astronomy, Wuhan University, Wuhan, 430072, China
}

\date{Accepted XXX. Received YYY; in original form ZZZ}

\pubyear{2023}

\begin{document}
\label{firstpage}
\pagerange{\pageref{firstpage}--\pageref{lastpage}}
\maketitle

\begin{abstract}
We present a spectral analysis of \textit{Insight}-HXMT observations of the low-mass X-ray binary 4U 1543-47 which locates in our Milky Way galaxy during the 2021 outburst. We focus on the observations in its soft state, and attempt to determine the spin of the black hole candidate through Thermal-Continuum Fitting (CF) method. The spin derived from CF method is highly dependent on black hole mass, distance and inclination angle of inner disk. In this article, we have adopted the preferred range of parameters: $M=9.4\pm 1 M_{\sun}$, D=$7.5{\pm}0.5$ kpc and $i=36.3^{+5.3}_{-3.4}$ degrees. We attain a moderate spin, $a=0.46\pm0.12$, which is consistent with previous results measured in the 2002 outburst. Besides, we notice the spectra show a wide blue shifted absorption feature between 8-10 keV which would originate from the highly ionized iron line. We try to fit the feature with {\em xstar} model and suggest that this feature may come from relativistic disk wind with a velocity of $v_{\rm wind}\sim 0.2c$. We attribute this relativistic disk wind to the super-Eddington accretion during the black hole outburst.
\end{abstract}




\section{Introduction}

Spin is a fundamental physical parameter of a black hole (BH), which has a significant effect on accretion process. The efficiency of accreting matter converts to radiation is sensitive to the black hole spin, and can vary up to an order of magnitude depending on whether the matter accretes onto a slowly or rapidly rotating black hole \citep{1972ApJ...178..347B}. The kinetic luminosity of relativistic jets produced by a black hole is probably tied to its spin state. However, ingredients necessary to generate jet are poorly understood. The state of accretion flow and the rotation of accretion disk will also influence the generation and properties of jet \citep{2020ARA&A..58..407D}.
Spin can help us understand stellar-mass black holes in Galactic X-ray binary systems. In such systems, the spin can be a window on the formation process of black holes which mostly formed from core collapse of massive stars \citep{2021ARA&A..59..117R}. The spin is commonly defined in terms of the dimensionless parameter $a_{*} \equiv a/M = cJ/GM^2$ , where $M$ and $J$ represent the black hole mass and angular momentum, respectively.

To measure the black hole spin, there are two prevailing methods: the continuum-fitting method and X-ray reflection method. Both methods are based on the assumption that the accretion disk is a geometrically thin, optically thick and innermost stable circular orbit (ISCO) effectively truncates the observable disk. The thermal continuum fitting (CF) method is based on the fact that the spin of a black hole influences the position of ISCO as well as the temperature of the inner disk. For higher spin, the smaller ISCO means that more binding energy is extracted from accretion matter and heat the inner disk to a higher temperature \citep{2021ARA&A..59..117R}. This method relies on accurate measurement to the system parameters of mass, distance, and inclination angle for the source. These parameters can be reliably measured via independent methods \citep{2009ApJ...706L.230M, 2001AJ....122.2668G, 1998ApJ...499..375O}. The CF method is applied more widely in black hole X-ray binaries than AGNs due to the difficulty in obtaining black hole masses and disk inclinations for AGNs. For X-ray reflection method, spin is measured from the gravitational redshift of spectral features (fluorescent lines, absorption edges, and recombination continua) close to the ISCO. This method does not require prior constraints on the mass and distance, and therefore can be applied to measure spin of both supermassive BHs in AGNs and stellar mass BHs in X-ray binaries. In addition, it can make an independent constraint on the disk inclination angle via its effect on the shape of spectral lines \citep{2014ApJ...793L..33M, 2020MNRAS.493.4409D}. In addition to above two methods, gravitational wave (GW) astronomy has opened a new window on black hole spin via the study of merging binary BHs. GW signatures of relativistic gravity, including spin, are “clean” in the sense that they are not subject to the complexities affecting our understanding of accretion flows. However, the imprints of spin on GWs can be subtle and, at the current level of sensitivity provided by the Laser Interferometer Gravitational Wave Observatory (LIGO), there are still only a small number of strong spin constraints \citep{2021ARA&A..59..117R}.

4U 1543-47 was first reported in 1971 and was classified as the black hole X-ray binary \citep{1972ApJ...174L..53M}. There are four outbursts before 2021 outburst, in 1971, 1984 \citep{1984PASJ...36..799K}, 1992 \citep{1992IAUC.5504....1H} and 2002 \citep{2004ApJ...610..378P}. Low-frequency quasi-periodic oscillations (QPOs) are found during the hard state of the 2002 outburst \citep{2004ApJ...610..378P}. The appearance of a compact jet during 2002 outburst is reported by \citet{2020MNRAS.495..182R}. Both continuum-fitting method and X-ray reflection method have been applied to estimate the spin of 4U 1543-47. \citet{2006ApJ...636L.113S} estimated a spin of $\sim0.75-0.85$ via continuum-fitting method. \citet{2009ApJ...697..900M} estimated a spin $a=0.3\pm0.1$ using a model combined by blurred reflection model \texttt{CDID} and continuum-fitting model \texttt{kerrbb}. \citet{2014ApJ...793L..33M} estimated a spin $a=0.43^{+0.22}_{-0.31}$ using \texttt{reflionx} and \texttt{kerrbb2}.
\citet{2020MNRAS.493.4409D} estimated a spin $0.67^{+0.15}_{-0.08}$ by using X-ray reflection method. These results are based on the data observed of the 2002 outburst, and independently determined black hole mass of $9.4\pm1.0 M_{\sun}$ and distance of $7.5\pm0.5$ kpc \citep{2004ApJ...610..378P}. \citet{2006ApJ...636L.113S} and \citet{2014ApJ...793L..33M} used the binary inclination $i=20.7\pm 1.5$ degree as disk inclination angle. \citet{2014ApJ...793L..33M} and \citet{2020MNRAS.493.4409D} set the disk inclination angle $i$ in reflection model free, and got $i\simeq 32^{+3}_{-4}$ and $36.3^{+5.3}_{-3.4}$ degree respectively. The results are slightly larger than the orbital inclination. When the spin vector of the BH is misaligned with the binary’s orbital angular momentum, the inclination of the inner disk may be misaligned with the orbital inclination since it aligns instead with the spin of the BH \citep{1975ApJ...195L..65B}. This also leads to the
Lense–Thirring precession of the inner hot flow \citep{2009MNRAS.397L.101I} and then produces the observed QPOs.

In this work, we aim to examine X-ray spectra observed by \textit{Insight}-HXMT during the 2021 outburst of 4U 1543-47 and to measure its spin in the soft state. We use a continuum-fitting method with the preferred range of parameters: $M=9.4\pm1 M_{\sun}$, $D=7.5\pm0.5$ kpc and $i=36.3^{+5.3}_{-3.4}$ degree. Observations, data selection and reduction procedures are described in Section 2. Analysis methods and results are presented in Section 3. We study the absorption feature in Section 4. We discuss the spectra and absorption feature fitting results in Section 5. Section 6 is summary of our work.

\section{Observations and data reduction}

\begin{figure}
	\includegraphics[width=\columnwidth]{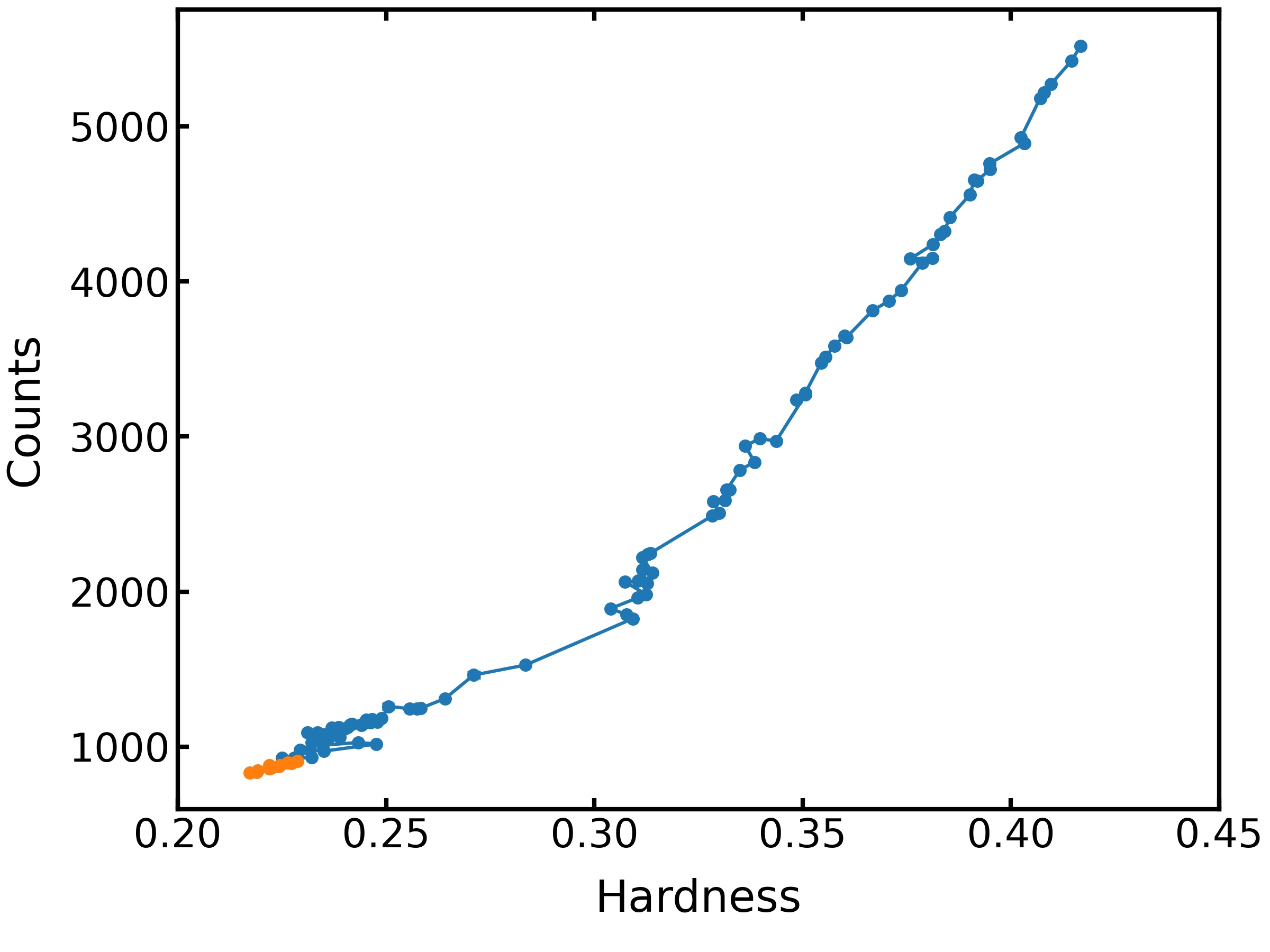}
    \caption{The evolutionary tracks in hardness-intensity diagram for \textit{Insight}-HXMT observations. The vertical axis presents the count rate in energy band 2–10 keV. The horizontal axis presents the hardness ratio (HR) defined as the ratio of count rate between 2 - 4 keV and 4 - 10 keV. The eleven orange dots represent the data we used to determine the spin of the black hole.}
    \label{figure1}
\end{figure}

The Hard X-ray Modulation Telescope \textit{Insight}-HXMT is a large X-ray astronomical satellite with a broad energy band in 1–250 keV. In order to fulfill the requirements of the broad band spectra and fast variability observations, three payloads are configured onboard \textit{Insight}-HXMT: High Energy X-ray telescope (HE) for 20–250 keV band \citep{liu2020high}, Medium Energy X-ray telescope (ME) for 5–30 keV band \citep{cao2020medium}, and Low Energy X-ray telescope (LE) 1–15 keV band \citep{chen2020low}. Light curves and spectra were extracted using \textit{Insight}-HXMT Data Analysis Software (HXMTDAS) v2.05 following the standard procedure (also see processing details described in \citealt{WANG20211,chen2021relation}). In the data screening procedure, we use tasks $he/me/lepical$ to remove spike events caused by electronic systems and $he/me/legtigen$ to select good time interval (GTI) when the pointing offset angle $< 0.04^\circ$; the pointing direction above earth $> 10^\circ$; the geomagnetic cut-off rigidity >8 GeV and the South Atlantic Anomaly (SAA) did not occur within 300 seconds. 

\begin{figure*}
	\includegraphics[width=2.\columnwidth]{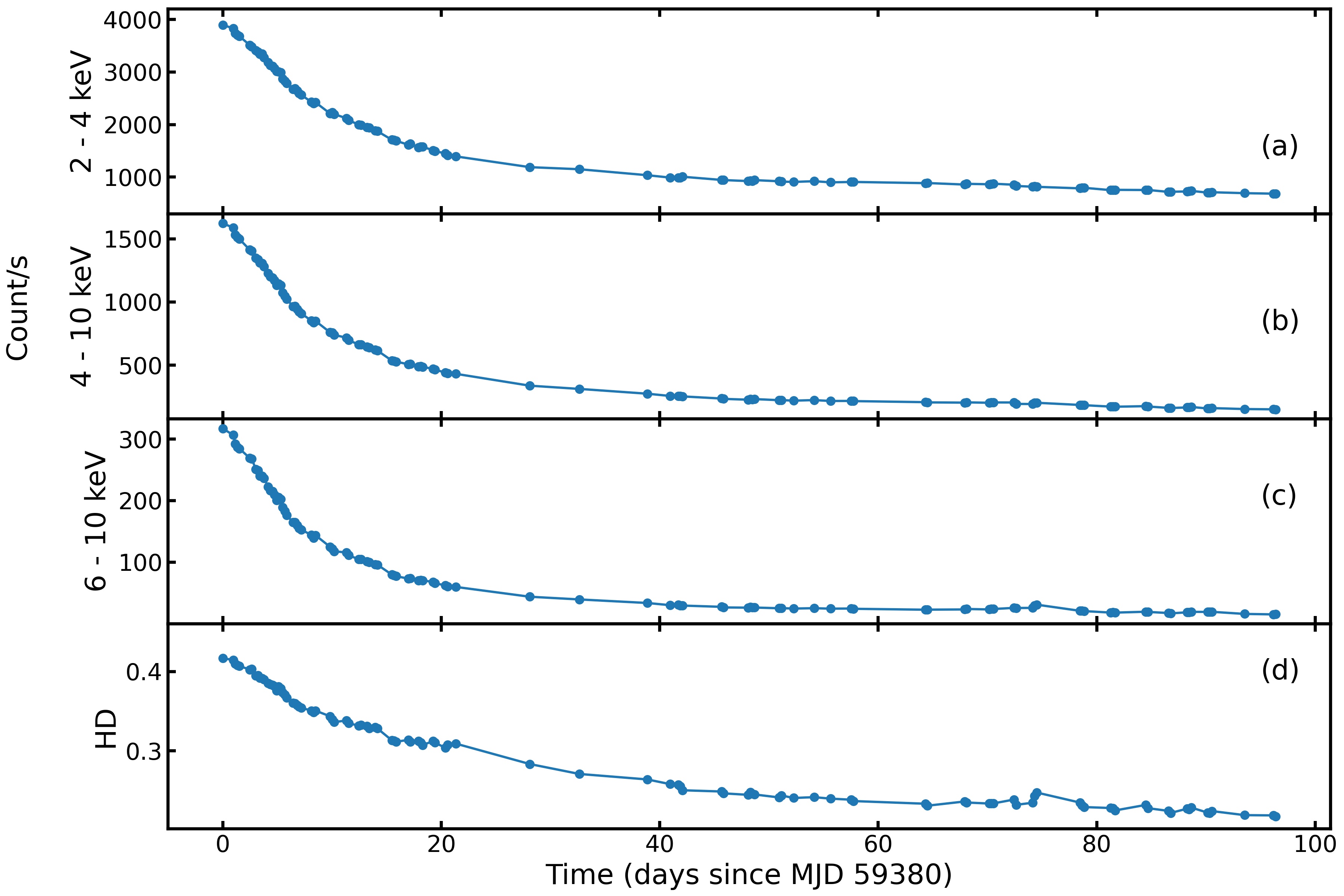}
    \caption{\textit{Insight}-HXMT observations light curve in the energy bands (a) 2-4 keV and (b) 4-10 keV (c) 6-10 keV. The hardness ratio (d) is defined as the ratio of count rate between 4-10 keV and 2-4 keV.}
    \label{figure2}
\end{figure*}
Our study is based on \textit{Insight}-HXMT observations of 2021 outburst of 4U 1543-47 from 15 June 2021 (MJD 59380) to 19 September 2021 (MJD 59476). We download all observations, plot the hardness-intensity diagram (HID) in Fig. \ref{figure1}. The BH candidate 4U 1543-47 showed a different behavior from other BH candidates (like Q-diagram) in the HID: 4U 1543-47 evolved from high luminosities with hard spectra to the low luminosities in soft state. In the softest state, we used the observations to perform the spectral analysis in the BH spin study. 

The light curve reveals that the source is very luminous in low energy band (2-4 keV) with a very high count rate (Fig. \ref{figure2}a), while the rate in the high energy band (6-10 keV, see Fig. \ref{figure2}c) is an order of magnitude lower. We selected eleven observations (MJD 59466 – MJD 59476) that were in the softest state during the observation period. In Table. \ref{table1}, we present the detailed information of these observations. During the 2021 outburst, 4U 1543-47 was extremely bright and very likely exceeded its Eddington luminosity \citep[see Fig. \ref{figure8} as well.]{2023arXiv230308837S, 2023MNRAS.520.4889P}. Spectra used to measure spin via CF method should not have the luminosity exceeding 30\% of the Eddington limit, since otherwise the accretion disk will become geometrically thick and may have nonzero torques at their inner boundary \citep{2006ApJ...652..518M, 2014ApJ...793L..33M}. The eleven selected observations have luminosities (1-20 keV) in range of $\sim 0.28-0.31 L_{\rm Edd}$. In this work, we analyzed the spectra using 2-10 keV for LE \citep{chen2020low}, 9-29 keV for ME \citep{cao2020medium} and 27-100 keV for HE \citep{liu2020high}.

\begin{table}
    \centering
    \caption{The list of HXMT observation IDs of the source 4U 1543-47 considered for the study, the hardness ratio (HR) defined as the count rate ratio between 4-10 keV and 2-4 keV.}
    \begin{tabular}{p{1cm} p{2.5cm} p{1.5cm} p{1.5cm}}
       \hline
       Num. & Observation ID & MJD & HR  \\
       \hline
       1 & P030402603701  &   59466  &     0.225  \\
       2 & P030402603702  &   59466  &     0.222  \\
       3 & P030402603703  &   59466  &     0.227  \\
       4 & P030402603801  &   59468  &     0.226  \\
       5 & P030402603802  &   59468  &     0.229  \\
       6 & P030402603803  &   59468  &     0.222  \\
       7 & P030402603901  &   59470  &     0.222  \\
       8 & P030402603902  &   59470  &     0.224  \\
       9 & P030402603903  &   59470  &     0.219  \\
       10 & P030402604001  &   59473  &    0.219  \\
       11 & P030402604101  &   59476  &    0.217  \\
       \hline
    \end{tabular}
    \label{table1}
\end{table}

\begin{figure}
	\includegraphics[width=\columnwidth]{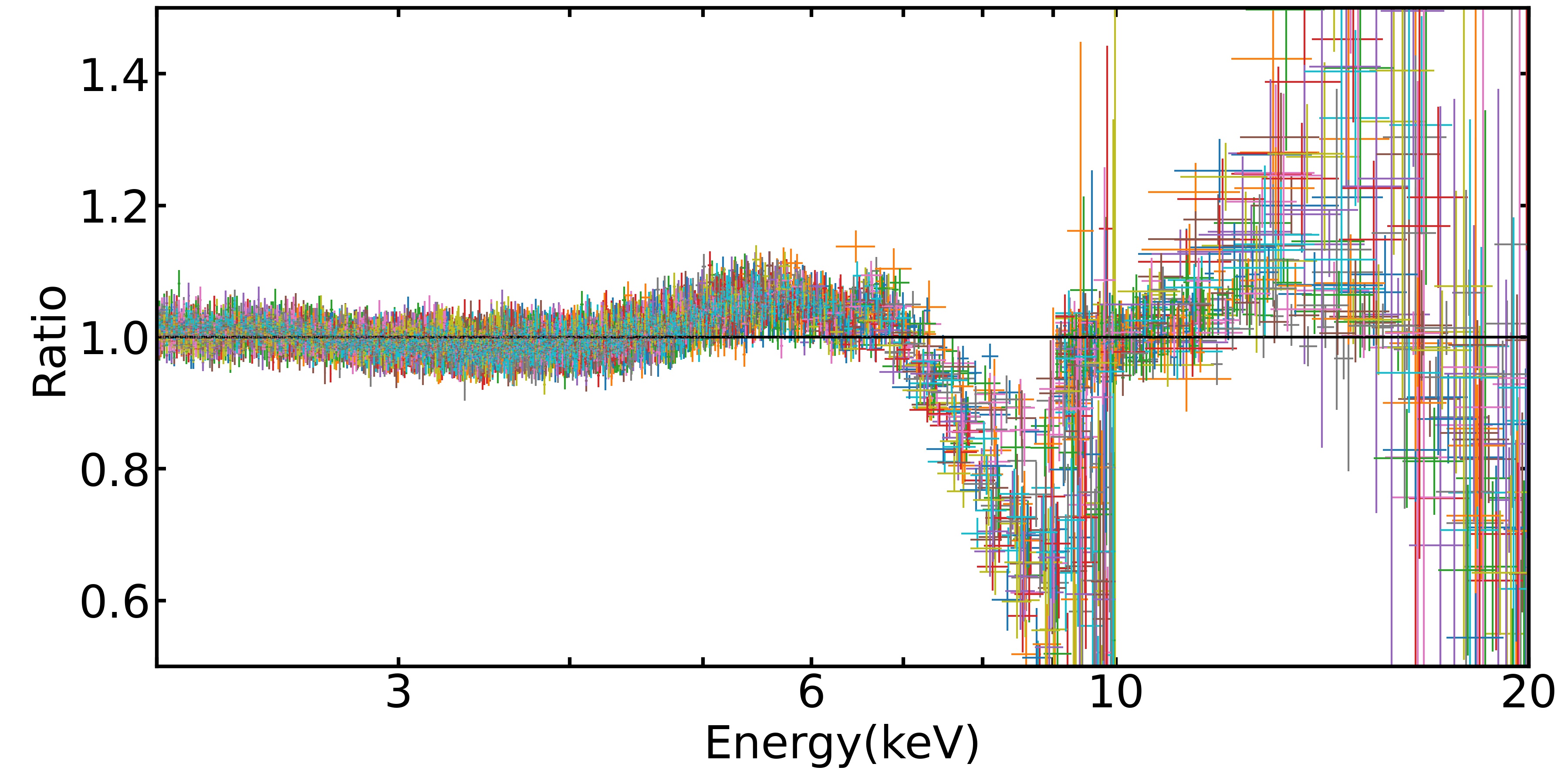}
    \caption{The ratios (model to data) of the fitting \textit{Insight}-HXMT observations with \texttt{constant$*$tbabs(diskbb+powerlaw)} model from 2 -- 20 keV. We use all observations during the whole observation period. There is a wide absorption feature between 8-10 keV in all observations from MJD 59380 -- 59476.}
    \label{figure3}
\end{figure}

\begin{figure*}
	\includegraphics[width=2\columnwidth]{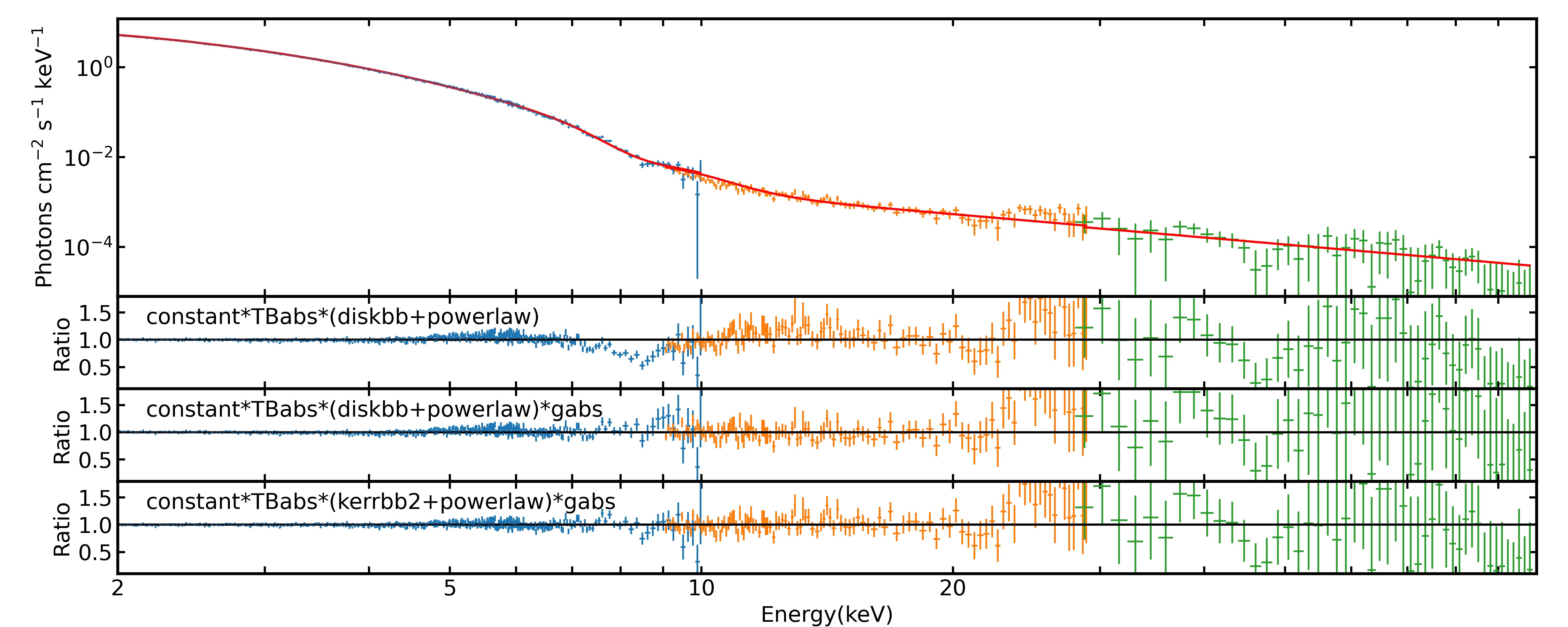}
    \caption{The result of fitting to observation 4 with model \texttt{tbabs(diskbb+powerlaw)}, \texttt{tbabs(diskbb+powerlaw)$*$gabs} and \texttt{tbabs(kerrbb2+powerlaw) $*$gabs}.}
    \label{figure3*}
\end{figure*}

\section{analysis and results}

We first fit the data with a phenomenological model, \texttt{constant$*$ tbabs(diskbb+powerlaw)}, in an attempt to check if there is any emission or absorption feature in the spectra. Fig. \ref{figure3} presents the data-to-model ratio results of all spectral fittings. We fit all observations during the whole observation period from the hard to soft state based on Insight-HXMT. There is a wide absorption feature between 8-10 keV in all observations including the soft state. This absorption feature is also reported by \citet{2023MNRAS.520.4889P}. Highly blue shifted absorption lines are observed in many AGNs \citep{2023A&A...670A.182M}. The absorption lines in NGC 4045 \citep{2022ApJ...929..138B} and WKK 4438 \citep{2018MNRAS.481..639J} are considered due to ultra-fast outflow. \citet{2023MNRAS.520.4889P} argued the absorption in 4U 1543-47 is coming from accretion disk-wind. The continuum-fitting model is mainly used to fit the thermal part of the spectrum. The absorption feature is in powerlaw part which would have little influence on the continuum-fitting result. Therefore, we use a simple Gaussian absorption model \texttt{gabs} to fit the absorption line when measuring the spin. We fit the spectra with model \texttt{constant$*$tbabs(kerrbb2+powerlaw)$*$gabs}. The model \texttt{kerrbb2} is a version of the \texttt{kerrbb} model \citep{2005ApJS..157..335L} that contains look-up tables generated using \texttt{bhspec} \citep{2006ApJS..164..530D}, and returns a self-consistent value of the spectral hardening factor ($f_{col}$) given various inputs of the spin and mass accretion rate. In Fig. \ref{figure3*}, we present the example spectrum from the observation ID P030402603801 (Num. 4 in Table \ref{table1}).

\begin{figure*}
	\includegraphics[width=1.6\columnwidth]{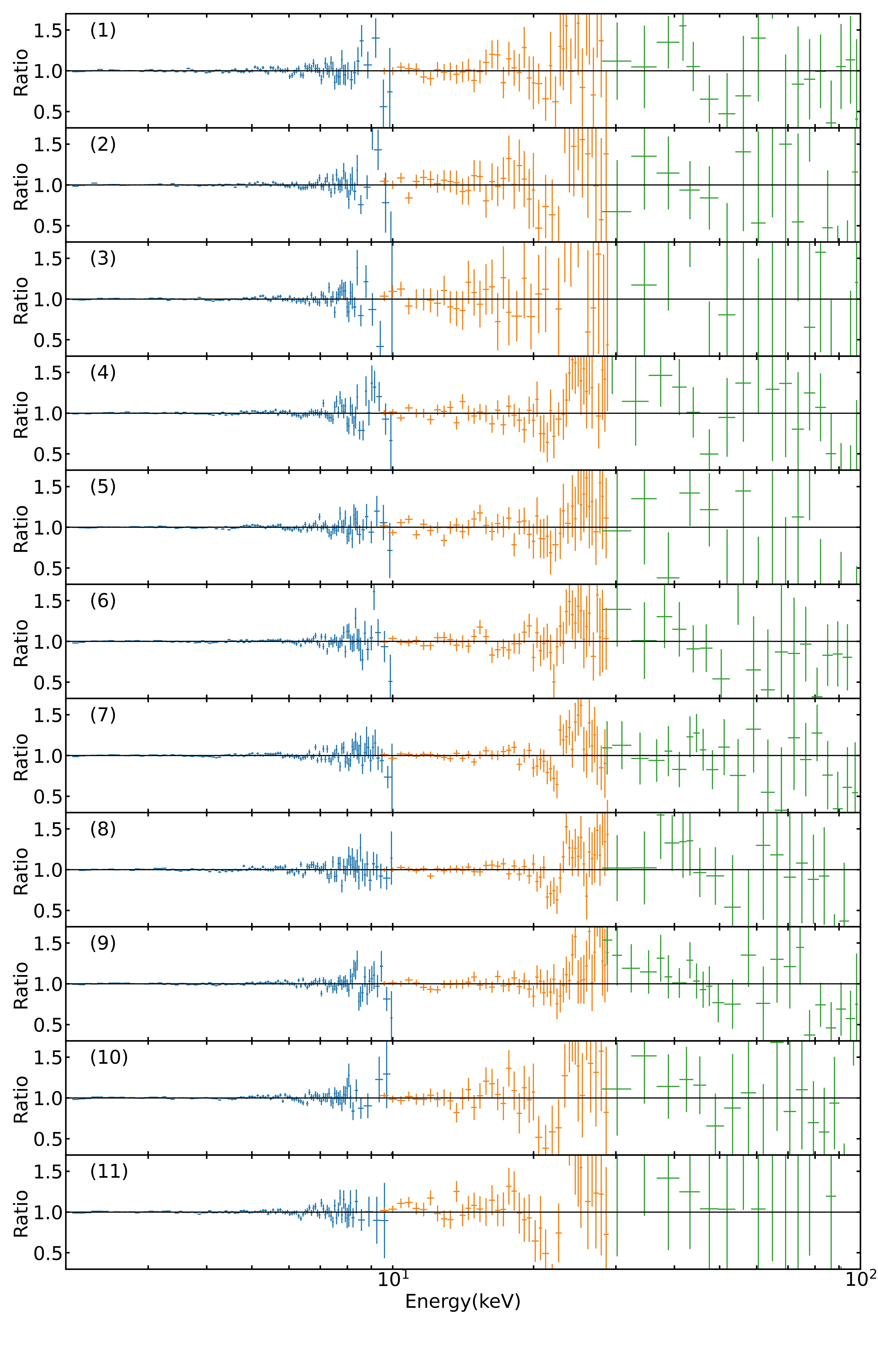}
    \caption{Making fit to eleven observations using the model: \texttt{constant$*$tbabs(kerrbb2+powerlaw)$*$gabs}. Data-to-model ratios are shown in the figure. Where marks in blue are from LE telescope and marks in orange are from ME telescope.}
    \label{figure4}
\end{figure*}

Usually, we expect that if we fix the normalization of LE to 1, and then the normalization of ME would also be 1. However, due to the effects of systematic errors, there are minor differences between the calibration of the two detectors \citep{2020JHEAp..27...64L}. During fitting process, the relative differences may change slightly. Model constant is used for coordinating calibration differences between the two detectors. We fixed the normalization of LE to 1, and the normalization of ME and HE are set free in range of 0.85 to 1.15. In model \texttt{kerrbb2}, we fix the mass, distance and accretion disk inclination angle to our favourite values. We adopt the black mass $M=9.4M_{\sun}$, the distance to the source $d=7.5$ kpc and the inclination of accretion disk $i=36.3$ degree which was estimated by \citet{2020MNRAS.493.4409D} using the \texttt{relxill} model. In model \texttt{gabs}, we set the initial value of line energy to be 9.0 keV, and we fit the centroid energy, width and intensity of the absorption line. The line energy distributes from $\sim 8.5-9.2$ keV with a line width of $\sim 1$ keV. Model \texttt{tbabs} is used to fit the galactic absorption \citep{2000ApJ...542..914W}. The galactic column density in \texttt{kerrbb2} and \texttt{tbabs} are linked and is fixed at $N_H=4.0\times10^{21}$ cm$^{-2}$ as in \citet{2004ApJ...610..378P}. The returning radiation flag and limb darkening flag are set to be 1. Since we fix the values of the mass, inclination and distance, the normalization is fixed to 1. The rest parameters are set to their default values. Table \ref{table2} lists the best-fit parameters. The returning self-consistent values of the spectral hardening factor $f_{col}$ are in range $1.72-1.74$. Fig. \ref{figure4} shows the data-to-model ratio of each observation. In addition, we also try to fit the non-thermal component with a model \texttt{simpl} which links the Compton-scattered flux with the thermal component. However, this has very little improvement comparing to \texttt{powerlaw} and the difference between spin results is within 3$\%$. We argue that very weak non-thermal component leads to the result. Hence, we keep using the model \texttt{powerlaw} in the following analysis.


\begin{table*}
\renewcommand{\arraystretch}{1.5}
	\caption{Best-fit parameters for spectra with the model \texttt{constant$*$tbabs(kerrbb2+powerlaw)$*$gabs} ($M=9.4M_{\sun}$ , $i=36.3$ deg and  $D=7.5$ kpc). The uncertainties listed are the maximum and minimum 90\% confidence limits found for each parameter}
	\label{table2}
	\begin{tabular}{c|c|c|c|c|c|c|c|c|c} 
		\hline
		Num.  &  \multicolumn{2}{c}{kerrbb2}  & \multicolumn{2}{c}{powerlaw}  & \multicolumn{3}{|c|}{gabs}  & $\chi^2$/d.o.f & Reduced $\chi^2_{\upsilon}$\\ \cline{2-8}
                  & $a_*$ & $\Dot{M}$ &  $\Gamma$ & Norm &  Line E & $\sigma$ & Strength & &\\
                  &   & $\times10^{18}$g cm$^{-2}$s$^{-1}$  & & & (keV)&  (keV)&(keV)   \\
		\hline
             1 & $0.504^{+0.007}_{-0.008}$ & $4.86^{+0.05}_{-0.05}$ & $1.59^{+0.16}_{-0.14}$ & $0.04_{-0.01}^{+0.03}$ & $9.04_{-0.22}^{+0.19}$ & $1.17_{-0.12}^{+0.12}$ & $2.49_{-0.41}^{+0.35}$ & 387/481 & 0.80  \\
             
             2 & $0.473^{+0.006}_{-0.006}$ & $5.07^{+0.04}_{-0.06}$ & $1.75^{+0.22}_{-0.19}$ & $0.05_{-0.02}^{+0.04}$ & $9.03_{-0.19}^{+0.19}$ & $1.27_{-0.09}^{+0.10}$ & $2.76_{-0.38}^{+0.43}$ & 460/481 & 0.96   \\
             
             3 & $0.471^{+0.007}_{-0.006}$ & $5.08^{+0.04}_{-0.07}$ & $1.91^{+0.40}_{-0.30}$ & $0.06_{-0.03}^{+0.11}$ & $9.33_{-0.18}^{+0.20}$ & $1.33_{-0.10}^{+0.11}$ & $3.41_{-0.40}^{+0.48}$ & 461/481 & 0.96   \\
             
             4 & $0.480^{+0.006}_{-0.006}$ & $5.04^{+0.04}_{-0.04}$ & $1.70^{+0.12}_{-0.11}$ & $0.09_{-0.03}^{+0.04}$ & $8.76_{-0.17}^{+0.18}$ & $1.10_{-0.09}^{+0.10}$ & $1.76_{-0.22}^{+0.26}$ & 464/481 & 0.97  \\
             
             5 & $0.481^{+0.006}_{-0.007}$ & $5.02^{+0.04}_{-0.05}$ & $2.07^{+0.16}_{-0.14}$ & $0.22_{-0.07}^{+0.13}$ & $9.20_{-0.20}^{+0.21}$ & $1.35_{-0.11}^{+0.12}$ & $2.43_{-0.34}^{+0.37}$ & 467/481 & 0.97   \\
             
             6 & $0.469^{+0.006}_{-0.005}$ & $5.16^{+0.03}_{-0.04}$ & $1.92^{+0.11}_{-0.10}$ & $0.17_{-0.04}^{+0.06}$ & $9.30_{-0.17}^{+0.17}$ & $1.35_{-0.10}^{+0.11}$ & $2.16_{-0.24}^{+0.25}$ & 502/481 & 1.04   \\
             
             7 & $0.437^{+0.007}_{-0.007}$ & $5.06^{+0.04}_{-0.05}$ & $2.21^{+0.08}_{-0.08}$ & $0.78_{-0.16}^{+0.20}$ & $8.84_{-0.22}^{+0.23}$ & $1.16_{-0.13}^{+0.14}$ & $0.95_{-0.15}^{+0.17}$ & 460/481 & 0.96   \\
             
             8 & $0.443^{+0.008}_{-0.008}$ & $5.02^{+0.05}_{-0.05}$ & $2.19^{+0.08}_{-0.08}$ & $0.70_{-0.15}^{+0.20}$ & $8.57_{-0.19}^{+0.22}$ & $0.95_{-0.13}^{+0.15}$ & $1.00_{-0.15}^{+0.17}$ & 483/481 & 0.98  \\
             
             9 & $0.451^{+0.005}_{-0.006}$ & $5.06^{+0.04}_{-0.03}$ & $2.08^{+0.08}_{-0.07}$ & $0.47_{-0.10}^{+0.13}$ & $8.99_{-0.21}^{+0.23}$ & $1.12_{-0.12}^{+0.13}$ & $1.26_{-0.18}^{+0.21}$ & 448/481 & 0.93  \\
             
             10 & $0.482^{+0.007}_{-0.006}$ & $4.87^{+0.04}_{-0.04}$ & $1.74^{+0.17}_{-0.15}$ & $0.05_{-0.02}^{+0.03}$ & $8.93_{-0.20}^{+0.20}$ & $1.20_{-0.10}^{+0.10}$ & $2.50_{-0.36}^{+0.40}$ & 464/481 & 0.96   \\
             
             11 & $0.481^{+0.007}_{-0.007}$ & $4.83^{+0.04}_{-0.04}$ & $1.91^{+0.24}_{-0.20}$ & $0.07_{-0.03}^{+0.06}$ & $9.20_{-0.17}^{+0.16}$ & $1.28_{-0.09}^{+0.09}$ & $3.17_{-0.40}^{+0.32}$ & 483/481 & 1.00   \\
		\hline
	\end{tabular}
\end{table*}

\begin{table*}
\renewcommand{\arraystretch}{1.5}
	\caption{Best-fit parameters for spectra with the model \texttt{constant$*$tbabs(kerrbb2+powerlaw)$*$gabs} ($M=9.4M_{\sun}$ , $i=36.3$ deg and  $D=7.5$ kpc) and $N_H$ in \texttt{tbabs} is set free. The uncertainties listed are the maximum and minimum 90\% confidence limits found for each parameter}
	\label{table3}
	\begin{tabular}{c|c|c|c|c|c|c|c|c|c|c} 
		\hline
		Num.  & TBabs & \multicolumn{2}{c}{kerrbb2}  & \multicolumn{2}{c}{powerlaw}  & \multicolumn{3}{|c|}{gabs}  & $\chi^2$/d.o.f & Reduced $\chi^2_{\upsilon}$\\ \cline{2-9}
              &  $N_H$  & $a_*$ & $\Dot{M}$ &  $\Gamma$ & Norm &  Line E & $\sigma$ & Strength & &\\
              &    &   & $\times10^{18}$g cm$^{-2}$s$^{-1}$  & & & (keV)&  (keV)&(keV)   \\
		\hline
             1 & $0.43_{-0.05}^{+0.05}$ & $0.490_{-0.024}^{+0.023}$ & $4.95_{-0.16}^{+0.17}$ & $1.60_{-0.14}^{+0.16}$ & $0.04_{-0.02}^{+0.03}$ & $9.02_{-0.22}^{+0.20}$ & $1.14_{-0.12}^{+0.13}$ & $2.39_{-0.42}^{+0.39}$ & 386/480 & 0.80  \\
             
             2 & $0.44_{-0.03}^{+0.03}$ & $0.450_{-0.019}^{+0.018}$ & $5.24_{-0.13}^{+0.13}$ & $1.77_{-0.19}^{+023}$ & $0.05_{-0.02}^{+0.05}$ & $8.99_{-0.19}^{+0.21}$ & $1.21_{-0.10}^{+0.10}$ & $2.58_{-0.38}^{+0.43}$ & 455/480 & 0.95   \\
             
             3 & $0.45_{-0.04}^{+0.03}$ & $0.446_{-0.019}^{+0.019}$ & $5.26_{-0.13}^{+0.14}$ & $1.93_{-0.31}^{+0.44}$ & $0.06_{-0.03}^{+0.14}$ & $9.30_{-0.18}^{+0.20}$ & $1.27_{-0.10}^{+0.11}$ & $3.23_{-0.41}^{+0.48}$ & 456/480 & 0.95  \\
             
             4 & $0.42_{-0.03}^{+0.03}$ & $0.471_{-0.017}^{+0.019}$ & $5.11_{-0.13}^{+0.12}$ & $1.72_{-0.12}^{+0.12}$ & $0.09_{-0.03}^{+0.04}$ & $8.74_{-0.17}^{+0.18}$ & $1.08_{-0.10}^{+0.10}$ & $1.71_{-0.23}^{+0.27}$ & 464/480 & 0.97   \\
             
             5 & $0.45_{-0.04}^{+0.03}$ & $0.457_{-0.019}^{+0.020}$ & $5.19_{-0.14}^{+0.12}$ & $2.08_{-0.15}^{+0.16}$ & $0.023_{-0.08}^{+0.14}$ & $9.16_{-0.21}^{+0.22}$ & $1.29_{-0.12}^{+0.13}$ & $2.27_{-0.35}^{+0.39}$ & 463/480 & 0.97   \\
             
             6 & $0.45_{-0.03}^{+0.03}$ & $0.445_{-0.015}^{+0.012}$ & $5.33_{-0.05}^{+0.11}$ & $1.93_{-0.10}^{+0.12}$ & $0.17_{-0.04}^{+0.07}$ & $9.27_{-0.20}^{+0.16}$ & $1.29_{-0.12}^{+0.06}$ & $2.04_{-0.30}^{+0.23}$ & 494/480 & 1.03   \\
             
             7 & $0.41_{-0.03}^{+0.03}$ & $0.429_{-0.016}^{+0.018}$ & $5.11_{-0.12}^{+0.11}$ & $2.22_{-0.08}^{+0.08}$ & $0.81_{-0.18}^{+0.23}$ & $8.84_{-0.22}^{+0.23}$ & $1.14_{-0.14}^{+0.15}$ & $0.93_{-0.15}^{+0.17}$ & 459/480 & 0.96   \\
             
             8 & $0.44_{-0.04}^{+0.04}$ & $0.421_{-0.018}^{+0.021}$ & $5.16_{-0.14}^{+0.12}$ & $2.22_{-0.09}^{+0.09}$ & $0.77_{-0.18}^{+0.24}$ & $8.53_{-0.18}^{+0.22}$ & $0.90_{-0.13}^{+0.15}$ & $0.93_{-0.15}^{+0.17}$ & 480/480 & 1.00  \\
             
             9 & $0.42_{-0.03}^{+0.03}$ & $0.440_{-0.015}^{+0.016}$ & $5.14_{-0.11}^{+0.11}$ & $2.09_{-0.08}^{+0.08}$ & $0.49_{-0.11}^{+0.15}$ & $8.97_{-0.22}^{+0.23}$ & $1.10_{-0.12}^{+0.13}$ & $1.22_{-0.18}^{+0.21}$ & 447/480 & 0.93  \\
             
             10 & $0.47_{-0.03}^{+0.03}$ & $0.444_{-0.018}^{+0.018}$ & $5.13_{-0.12}^{+0.13}$ & $1.77_{-0.15}^{+0.17}$ & $0.05_{-0.02}^{+0.03}$ & $8.85_{-0.20}^{+0.20}$ & $1.12_{-0.10}^{+0.11}$ & $2.22_{-0.36}^{+0.41}$ & 451/480 & 0.94   \\
             
             11 & $0.46_{-0.04}^{+0.03}$ & $0.453_{-0.020}^{+0.019}$ & $5.02_{-0.13}^{+0.13}$ & $1.92_{-0.21}^{+0.24}$ & $0.07_{-0.03}^{+0.07}$ & $9.16_{-0.10}^{+0.10}$ & $1.21_{-0.10}^{+0.10}$ & $2.99_{-0.41}^{+0.33}$ & 476/480 & 0.99   \\
		\hline
	\end{tabular}
\end{table*}

Besides, we also attempt to fit the spectra under a free galactic column density and the best-fit parameters are listed in Table \ref{table3}. The column density is slightly higher than $4.0\times10^{21}$ cm$^{-2}$, and measured spins are close to the results obtained from fixed column density. The spin parameter obtained in fixed $N_H$ and free $N_H$ are in range $0.4-0.5$, which is consistent with the result in \citet{2009ApJ...697..900M} and \citet{2014ApJ...793L..33M}. However, the effective low energy of \textit{Insight}-HXMT is limited at 2 keV due to the worse calibration below 2keV. The data cannot constrain the $N_H$ well. Therefore, we fit the spectra with $N_H$ fixed at $4.0\times10^{21}$ cm$^{-2}$ in the following work.

\begin{figure*}
	\includegraphics[width=2\columnwidth]{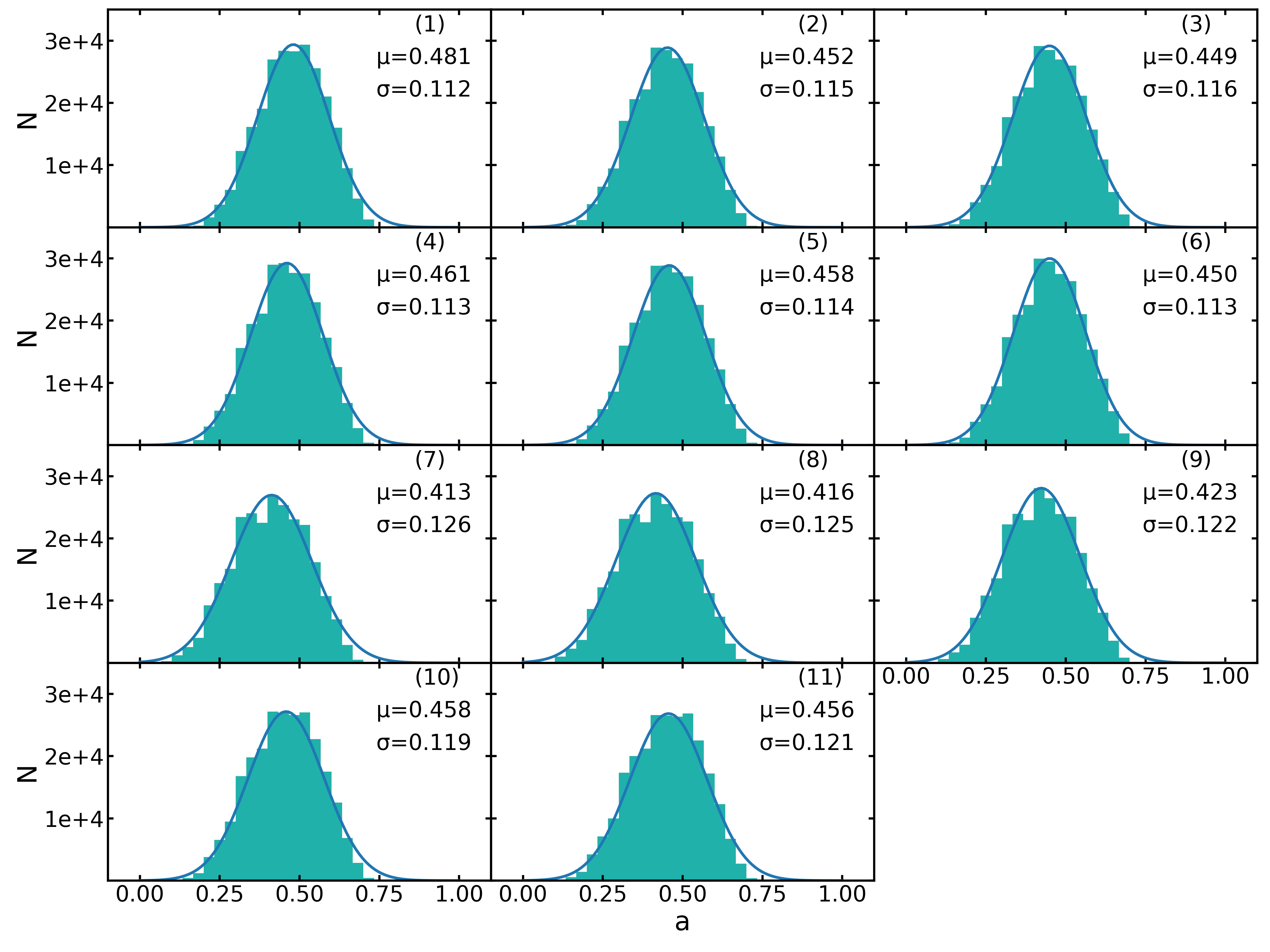}
    \caption{Histograms of $a_*$ calculated via the Monte Carlo analysis for each spectrum. A Gaussian function is used to fit the data, the fitted centroid value $\mu$ and $\sigma$ is shown in each sub-figure.}
    \label{figure5}
\end{figure*}

\begin{figure*}
	\includegraphics[width=1.2\columnwidth]{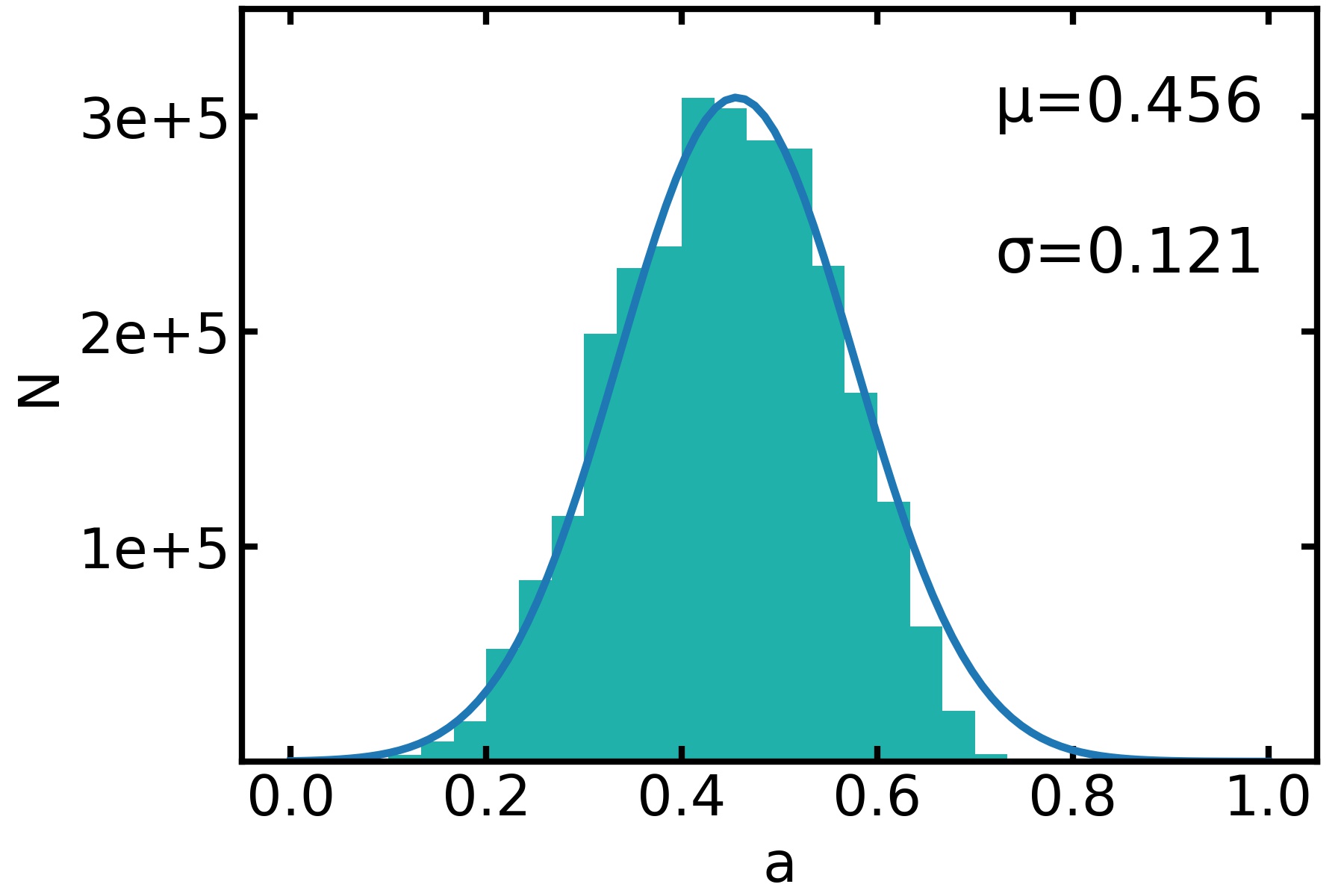}
    \caption{Summed histogram of $a_*$ for all 11 spectra.}
    \label{figure6}
\end{figure*}

To allow us estimate systematic uncertainties in the fitting that are produced by uncertainties in the mass, distance and accretion disk inclination angle, we adapt the Monte Carlo (MC) simulation method used in \citet{2021ApJ...916..108Z}. We fit the spectra for 50 evenly spaced values of mass in range $8-11 M_{\sun}$, 50 evenly spaced values of distance in range $6-9$ kpc, and 100 evenly spaced values of accretion disk inclination angle in range $20-42$ degrees. The results are presented in Fig. \ref{figure5} and Fig. \ref{figure6}. We use a Gaussian function to fit the summed spin results and attain a spin $a=0.46 \pm 0.12$ (1$\sigma$). In the model \texttt{kerrbb2}, the spin decreases as the inclination or distance increases, and increases as the mass increases (also see Fig. 5 in \citealt{2021ApJ...916..108Z}).

\section{Absorption features}

\begin{figure}
	\includegraphics[width=1.\columnwidth]{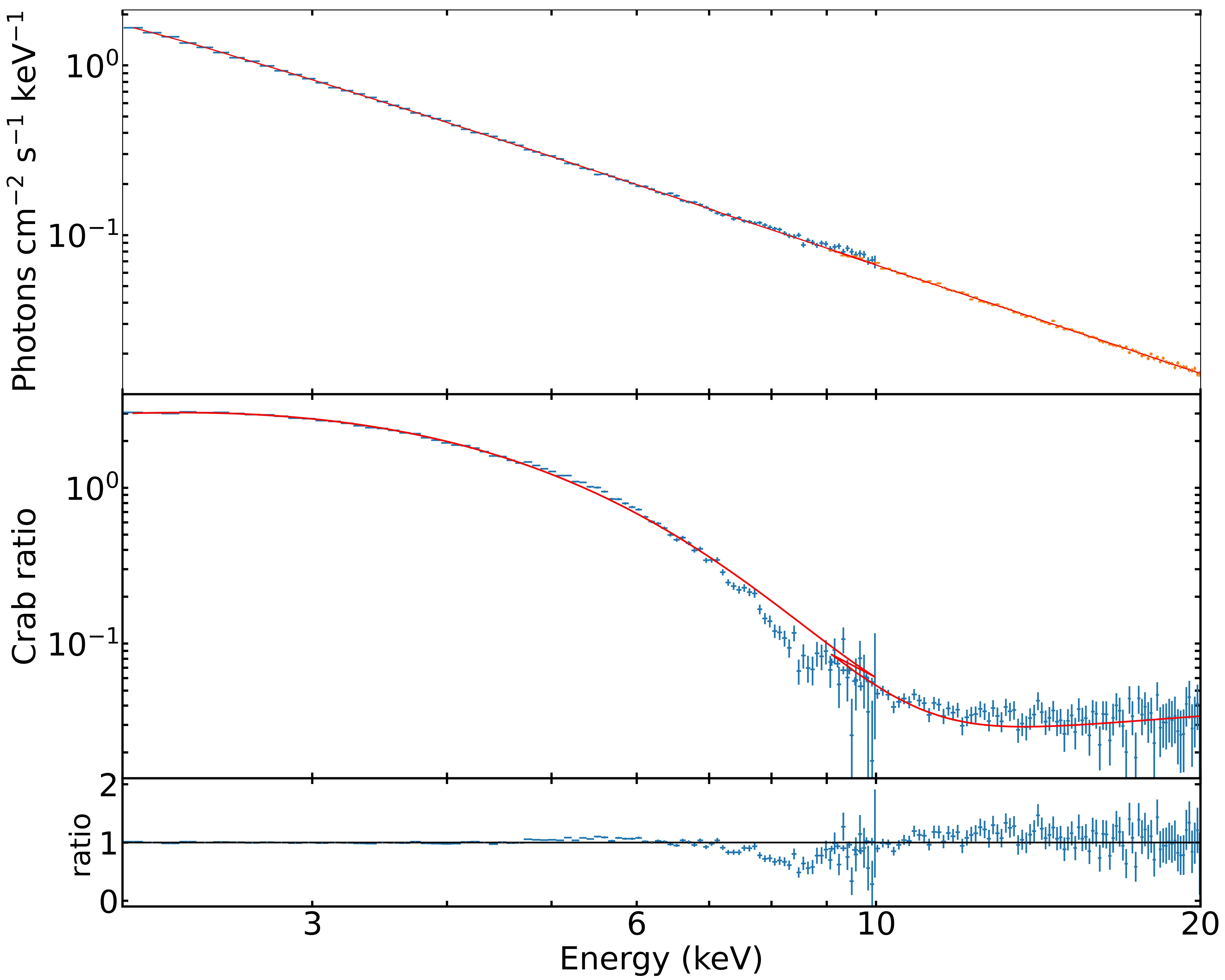}
    \caption{Upper panel: fitting result of the Crab spectrum (obs ID: P0402-34900102, MJD 59343) from Insight-HXMT. We use \texttt{powerlaw} model to fit, which gives $\Gamma=2.13\pm 0.01$, there are no any structures in the Crab spectrum. Middle and bottom panels: the ratio spectrum of the source for ObsID P030402603801 to Crab data. We use model \texttt{constant$*$tbabs(diskbb+powerlaw)} to fit the data. An absorption feature still exists between 8-10 keV. }
    \label{figure7}
\end{figure}

When fitting the spectra of the source, we have noticed a wide absorption feature between 8-10 keV during the whole observation period (see Fig. \ref{figure3}). In last section, we used a phenomenological model \texttt{gabs} when estimating the spin, here we would study the physical parameters of the absorption feature. In this section, we analyzed the spectra using 2-10 keV for LE , 9-20 keV for ME. Before doing that, we checked if the absorption feature is coming from instrumental or calibration errors. We analyzed the spectrum of the Crab in the nearest HXMT observation (obs ID: P040234900102, MJD 59343), and then derived the Crab ratio spectrum via dividing the BH source data (ObsID P030402603801) by Crab data. Fig. \ref{figure7} shows the result, where a wide absorption line still exists between 8-10 keV. The spectrum is fitted with model \texttt{constant$*$tbabs(diskbb+powerlaw)}. In general, the absorption features in the spectrum can be due to multiple reasons like the presence of obscuring cloud in the line-of-sight, occultation due to the companion star, disk reflection, strong accretion disk-wind and/or the stellar wind from the companion. Considering the absorption feature is highly blue-shifted, this absorption is probably coming from outflow or disk-wind. Besides, the parameters of \texttt{gabs} indicate that the absorptive matter has a relativistic turbulent velocity. We generate a photoionized absorption model \texttt{xstar} which is used to fit the parameters of disk wind. We use a MPI version of \texttt{xstar2xspec}, provided by \citet{2018PASP..130b4501D} in order to speed up the process, to generate a custom absorption model to be used in XSPEC. The model is calculated with a fixed turbulent velocity of 30,000 km s$^{-1}$, which is calculated from Gaussian line width. Free parameters are the ionization of the plasma log$(\xi)$, the column density $N_H$, the iron abundance Z$_{\textrm Fe}$ and the redshift z.

\begin{figure*}
	\includegraphics[width=2\columnwidth]{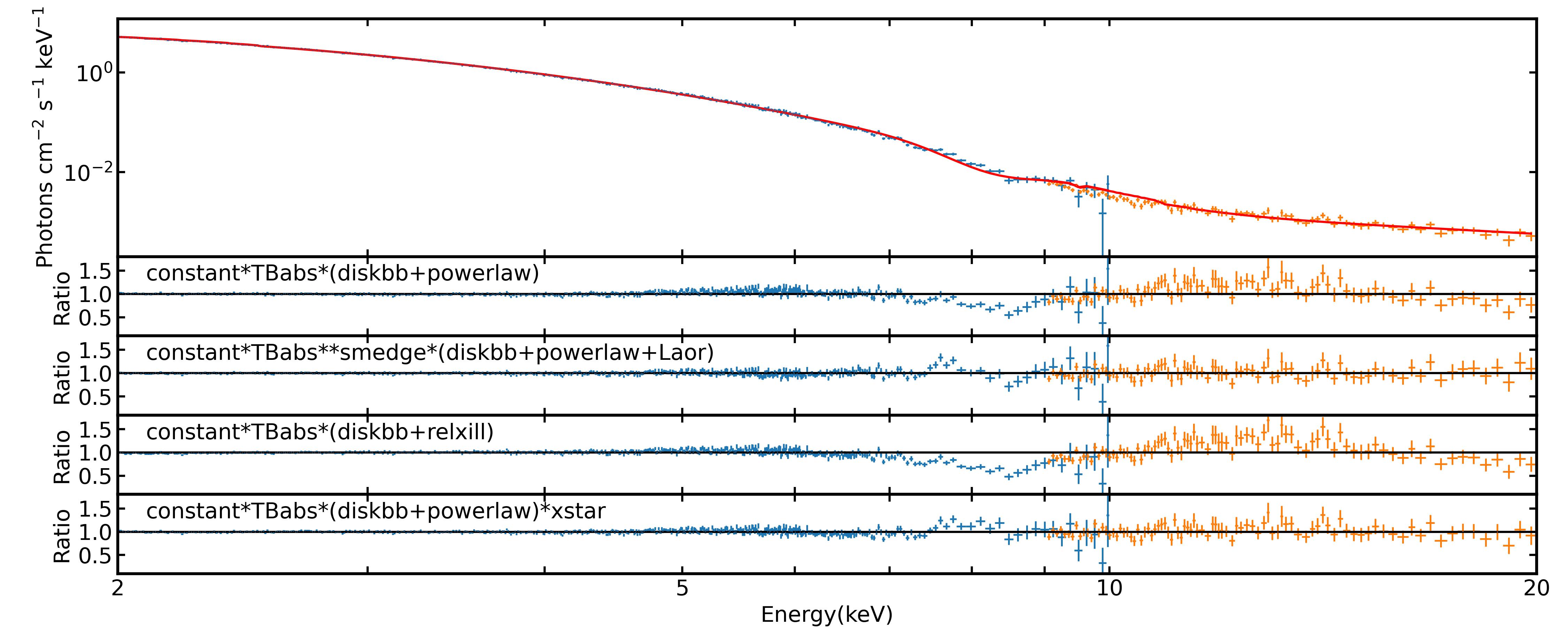}
    \caption{The results of fitting to the data of ObsID P030402603801 with the models \texttt{tbabs(diskbb+powerlaw)}, \texttt{tbabs$*$smedge(diskbb+powerlaw+ gaussian)}, \texttt{tbabs(kerrbb2+relxill)} and \texttt{tbabs(kerrbb2+powerlaw)$*${xstar}}.}
    \label{figure9}
\end{figure*}

Disk reflection and partial absorption are two possible mechanisms to generate broad absorption structure. In this section, we exam both reflection and disk wind absorption models. We have tried four models to fit the spectrum of the BH candidate. Results are shown in Fig. \ref{figure9} (as the example from ObsID P030402603801). The first model is \texttt{constant$*$tbabs(diskbb+powerlaw)}. This returns $\chi^2_{\upsilon}=1.37$. The data-to-model ratio plot shows an absorption feature. The absorption feature looks like the profile of a relativistically-broadened Fe-K line and associated smeared Fe-K edge. Therefore, we use phenomenological model \texttt{tbabs$*$smedge(diskbb+powerlaw+Laor)} to model the reflection feature. We set the line energy of model \texttt{Laor} free within 6.4-7.0 keV, and inclination to be 36.3$^\circ$. The threshold energy in \texttt{smedge} is set within 7-10 keV and the rest parameters are set as default. This gives $\chi^2_{\upsilon}\sim 1.0$, and fits the absorption feature well. However, we notice the normalization of model \texttt{Laor} is smaller than $10^{-3}$ which is a weak component.

In the course of the reflection, an iron fluorescence K-line is necessarily emitted at 6.4 keV, accompanying the iron K-ionization \citep{1994PASJ...46..375E}. The weak Fe K-line emission indicates the reflection component is weak. We then try the model \texttt{tbabs(kerrbb2+relxill)}. The inclination and spin in \texttt{relxill} and \texttt{kerrbb2} are linked. This model gives $\chi^2_{\upsilon}=1.74$. The data-to-model ratio plot still shows the absorption feature. The \texttt{relxill} does not improve the quality of fit, and the absorption is not well fitted. This is probably due to the weak Fe K-line emission. Besides, \citet{2023MNRAS.520.4889P} has to multiple the model \texttt{gabs} to fit the spectrum with model \texttt{relxilllp}. These indicate that the absorption feature is likely not due to the reflection features of the disk. Finally, we fit with \texttt{constant$*$tbabs(kerrbb2+powerlaw)$*$xstar}. The model spectrum based on the \texttt{xstar} model plus a simple powerlaw continuum is shown in Fig. \ref{figure10}. This improves the quality of fit significantly to $\chi^2_{\upsilon}=1.02$, and the absorption feature is well fitted. The best-fitting parameters of \texttt{xstar} are shown in Table \ref{table4}. 

We get a spin $a=0.40^{+0.08}_{-0.01}$, which is in consistent with our previous result $a=0.46\pm 0.12$. The result shows the absorptive matter has a column density $N_H=1.40^{+0.05}_{-0.02}\times10^{23}$ cm$^{-2}$, an ionization $\textrm{log}(\xi)=4.16^{+0.11}_{-0.02}$, a redshift $z=-0.186^{+0.002}_{-0.003}$ and an iron abundance $\textrm{Z}_{\textrm Fe}=3.18^{+0.02}_{-0.05}\textrm{Z}_{\sun}$. The iron abundance is probably degenerate with the column density. The redshift is corresponding to a line-of-sight velocity $v\sim0.2c$, and the turbulent velocity is comparable to it. Relativistic jet usually has a good collimation and will not generate a wide absorption line. Thus, we argue that this absorption is due to disk wind.

Numerical simulation shows a non-rotating black hole can generate relativistic disk wind when disk luminosity over Eddington luminosity \citep{2019PASJ...71...70T}. The disk wind will be accelerated as flowing outward and the terminal velocity can reach up to 0.4c. If rotation of disk and gas pressure is considered, the relativistic disk wind can be generated even if disk luminosity smaller than Eddington luminosity \citep{2019PASJ...71...70T}. For an extreme Kerr black hole (a=1), the terminal speeds of disk wind can reach 0.4c when the normalized accretion rate is $\Dot{m}=2$. The normalized accretion rate is defined as $\Dot{m}=(1/\eta)(L_d/L_E)$, where $\eta$ is energy transfer efficiency, $L_d$ is disk luminosity and $L_E$ is Eddington luminosity. The terminal speeds are different if the winds have different initial starting radius, which may lead to the velocity turbulence \citep{2001PASJ...53..285H}. 
The 2021 outburst of 4U 1543-47 was extremely bright and very likely exceeded its Eddington luminosity \citep{2023arXiv230308837S, 2023MNRAS.520.4889P}. The absorption feature is present thorough the observation, which indicates the wind would need to be continuously supplied. Eddington ratio of most observations $L/L_E\textless 1$, which are not likely to generate relativistic disk winds if ignoring the influence of rotation of the disk and gas pressure. However, when scattering and absorption are taken into consideration, the observed source luminosity is probably smaller than luminosity of disk accretion \citep{2021MNRAS.502.5797Y}. In this circumstance, the BH with an observed luminosity less than Eddington luminosity may still process the super-Eddington accretion and generate relativistic winds. Besides, GR-MHD simulations show that relativistic outflow can launch from accretion disk \citep{2019ApJ...882....2V}. 




In addition, we try to fit all observations with model \texttt{constant$*$ tbabs(diskbb+powerlaw)$*$xstar} to study the evolution of the disk wind. The source has a luminosity exceeding 30\% of Eddington luminosity during its early stage, which is not appropriate to apply \texttt{kerrbb2} model. Thus, we select a simple \texttt{diskbb} model. Due the degeneration between column density and iron abundance, the parameters can not be constrained well if all are set free. Therefore, we fix iron abundance to the best-fitting result of the example observation. The results of well constrained observations are shown in Fig. \ref{figure11}.

\section{Discussion}

In this article, we have performed a detailed spectral analysis of \textit{Insight}-HXMT data of the 2021 outburst of the black hole X-ray binary 4U 1543-47. The spectra of this outburst evolved from the hard state to the soft state. Considering the source was very soft in the later stage with the luminosity near $\sim 30\%$ Eddington luminosity, we selected continuum-fitting model to measure the spin parameter. By conducting a Monte Carlo (MC) simulation, we finally obtain the spin $a=0.46\pm 0.12$ (1$\sigma$). Our result is consistent with the measurement of \citet{2009ApJ...697..900M}, a=$0.3\pm0.1$, and \citet{2014ApJ...793L..33M}, a=$0.43^{+0.22}_{-0.31}$, which are measured using the 2002 outburst data. This indicates that the spin of 4U 1543-47 has a steady spin in the last twenty years. However, our result is lower than that obtained by \citet{2020MNRAS.493.4409D} using a reflection model, $a=0.67^{+0.15}_{-0.08}$. There is a hint of systematic bias between the methods, where reflection model usually gives a higher spin than continuum-fitting model \citep{2021ARA&A..59..117R}. But this hint is by no means concrete and out of a small number of comparison sources. There are already exceptions such as GRS 1915+105 and Cyg X-1 with extreme high continuum-fitting spins which have had less extreme reflection spin measurements.

The inclination $i$ in continuum-fitting model is supposed to be the inclination angle of the inner disk, which is hard to estimate in practice. The inclination is a crucial input model parameter and has an obvious degeneracy with the spin. Usually, the strategy is to assume it’s same as the orbital inclination or jet inclination. Some previous works on fitting the reflection component reported a small misalignment between the inner accretion disk and the binary orbital plane \citep{2010ApJ...719L..79F,2016ApJ...826...87W}, which may be due to the spin vector of the black hole misaligned with the binary’s orbital angular momentum \citep{1975ApJ...195L..65B}. In this work, we adopt the inner disk inclination estimated by \citet{2020MNRAS.493.4409D} using reflection model, which is a little bit larger than the binary inclination $i=20.7\pm1.5$ deg \citep{2003IAUS..212..365O}. More consistent and accurate dynamical parameters are required for the detailed spin measurements in the future.

\begin{figure}
	\includegraphics[width=1.\columnwidth]{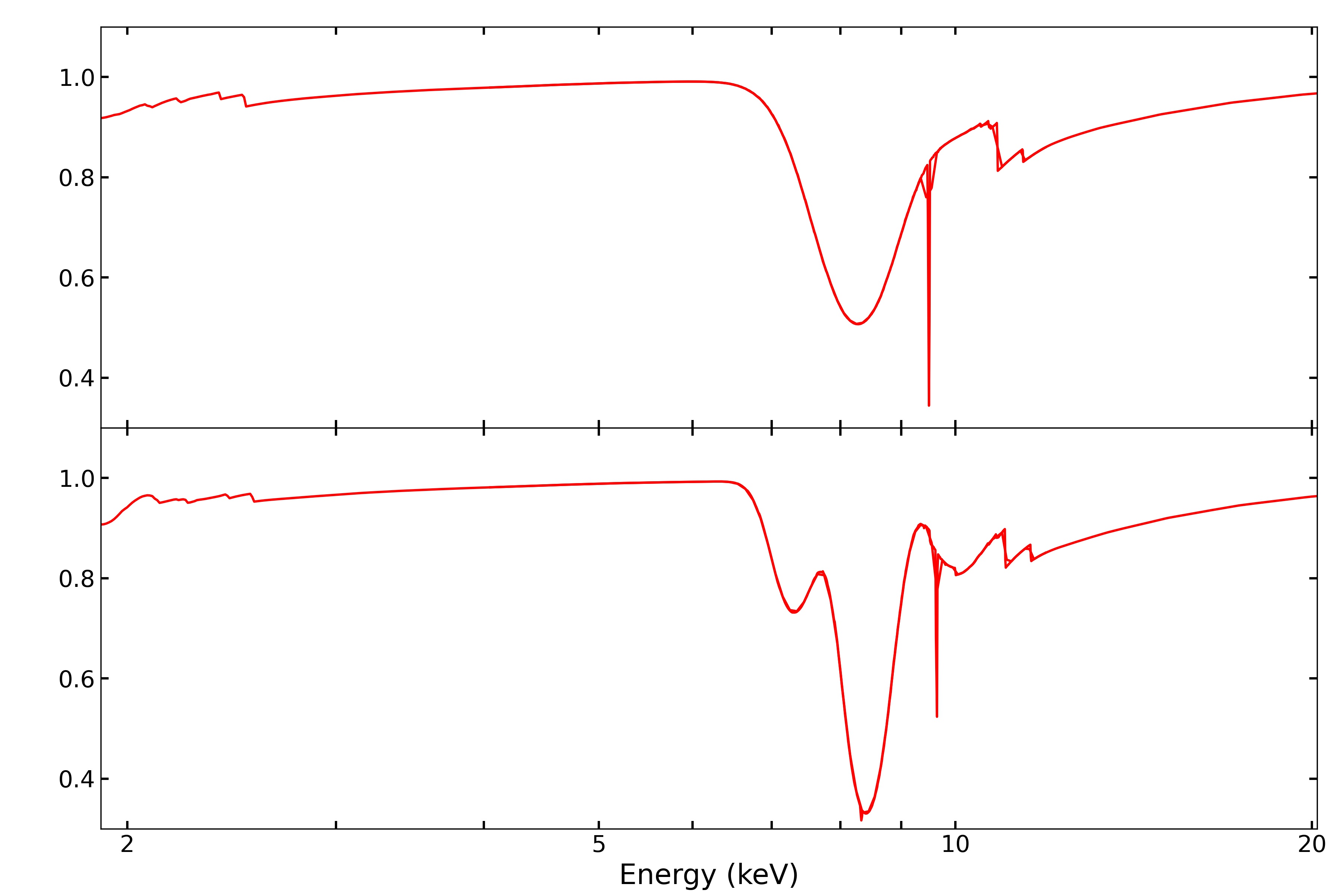}
    \caption{The \texttt{xstar} model applied to a simple powerlaw continuum. Upper panel: single absorption line. Lower panel: two absotption lines.}
    \label{figure10}
\end{figure}

We reported the absorption feature around 8--10 keV in all observations, which suggested the fast disk winds of an iron abundance $\textrm{Z}_{\textrm Fe}\sim  3\textrm{Z}_{\sun}$. Previous works report that the accretion disk is iron over-abundance, e.g., \citet{2020MNRAS.493.4409D} reports a value $\textrm{Z}_{\textrm Fe}=5.05^{+1.21}_{-0.26}\textrm{Z}_{\sun}$, \citet{2023MNRAS.520.4889P} gives a value $\textrm{Z}_{\textrm Fe}>5\textrm{Z}_{\sun}$. Highly blueshifted absorption lines are observed in many AGNs \citep{2023A&A...670A.182M}, but seldom in X-ray binaries. An 8.56 keV absorption line was observed in NGC 4045 \citep{2022ApJ...929..138B}, an absorption line feature around 10 keV was observed in WKK 4438 \citep{2018MNRAS.481..639J}. The width of these observed absorption lines are narrow and not larger than 1 keV. \citet{2018ApJ...854L...8R} reported two wide absorption lines in luminous quasar PDS 456. The turbulent velocity of this source reached 15,000 km s$^{-1}$. The 4U 1543-47 has a wide absorption feature, $\sigma>1$ keV, which indicates a high turbulent velocity $v_t\sim 0.1c$. This is comparable to the line-of-sight velocity of absorptive matter and has never been observed in X-ray binary systems before. The winds coming from different initial starting radius may lead to observed large turbulent velocity. Another possible explanation is that the absorption feature comes from the combination of multiple lines with different blue-shifts. We try to fit the observation data (ObsID P030402603801) with two \texttt{xstar} models which are generated with a turbulent velocity of $\sim$ 14,000 km s$^{-1}$. We freeze iron abundance to be 3.18Z$_{\sun}$ and link the ionization log$(\xi)$ of two models. This returns $\chi^2_{\upsilon}=0.99$, with a spin value of $a_*=0.41\pm 0.02$. The best-fitting parameters are listed in Table \ref{table5}. The two absorption models could fit the spectrum well. In our model, we assume both winds are highly ionized and both lines are Fe XXVI Ly$\alpha$ line (6.97 keV). The model spectrum based on two $xstar$ models plus a simple powerlaw continuum is shown in the bottom panel of Fig. \ref{figure10}. The observed absorption line can be composed of two absorption lines with a smaller turbulent velocity. The two lines have different blue-shifts, which indicates that they are two distinct winds. One possible explanation is that two disk winds generated in this outburst. The former one decelerates and dissipates earlier, which leads to the shallow absorption with a blue-shift of $z\sim -0.071$, the other line has a blue-shift of $z\sim -0.2$. However, constrained by the spectral resolution, we cannot constrain two absorption lines in other observations.

\begin{figure}
	\includegraphics[width=1.\columnwidth]{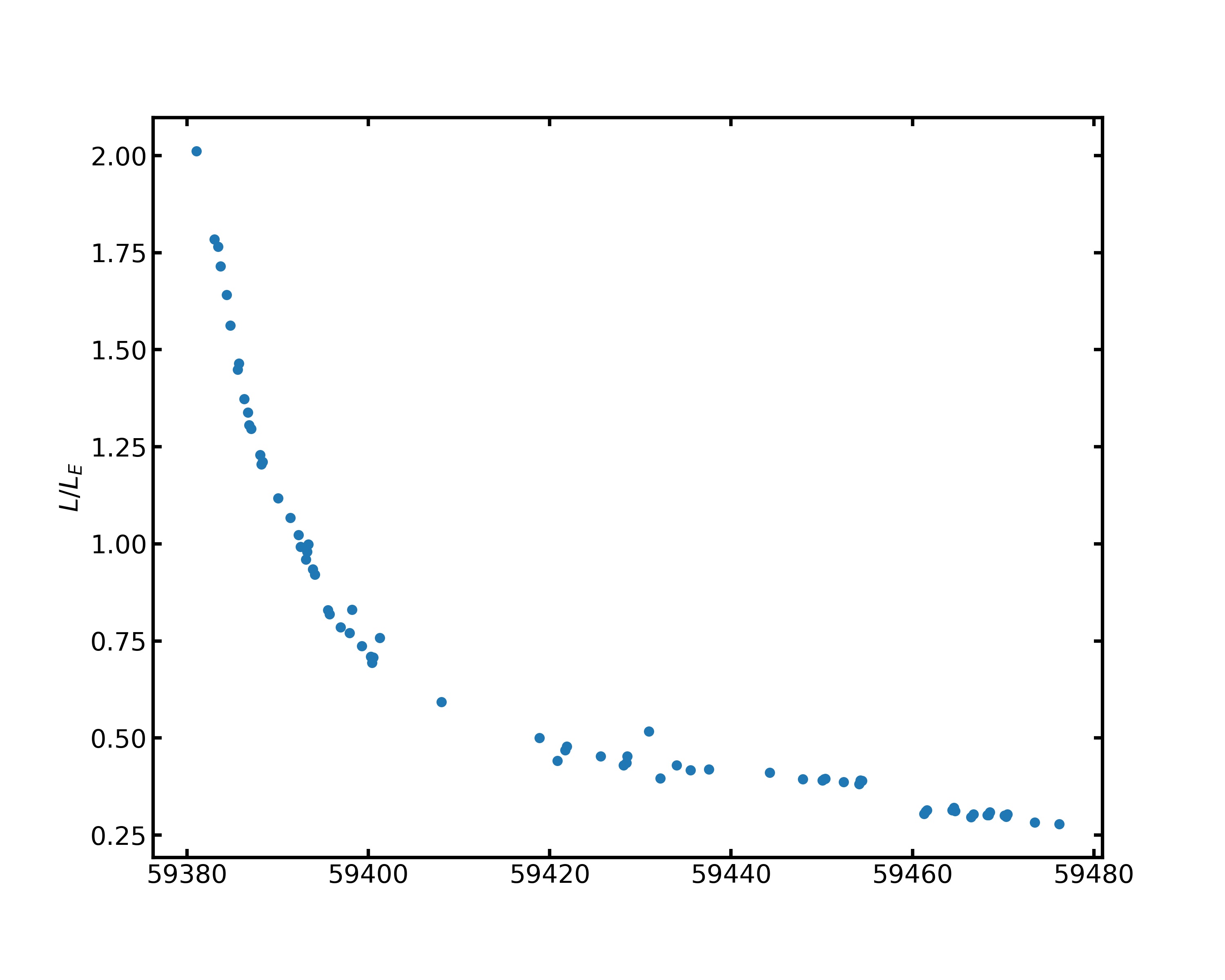}
    \caption{The ratio between source luminosity (1-20 keV) to its Eddington luminosity. Source luminosity is calculated from model, Eddington luminosity is calculated from our preferred BH parameters.}
    \label{figure8}
\end{figure}

\begin{figure*}
	\includegraphics[width=2.\columnwidth]{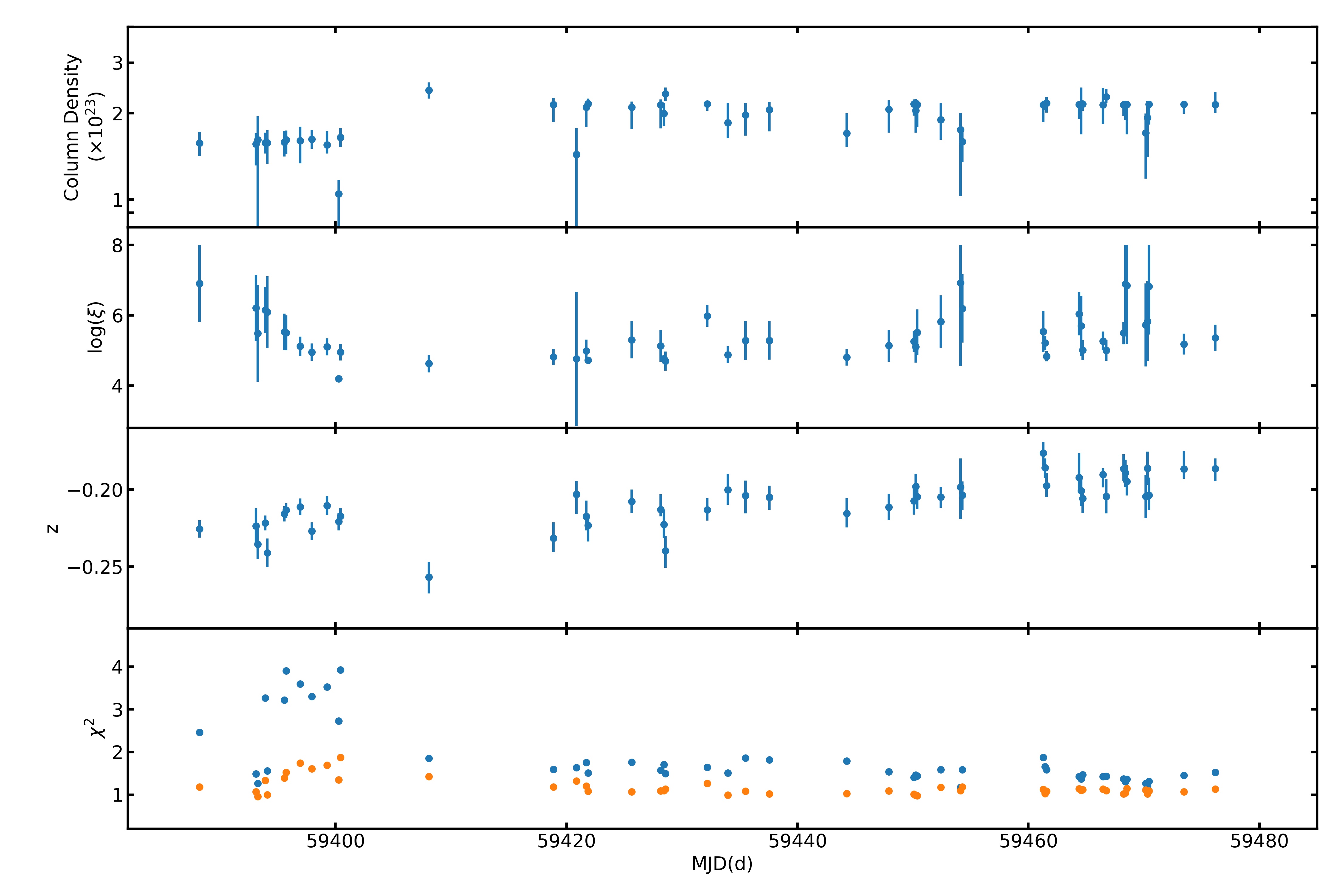}
    \caption{Evolution of parameters in model \texttt{xstar} with time. The source has a luminosity exceeding Eddington luminosity during its early stage, which is not appropriate to apply \texttt{kerrbb2} model. Thus, we select a simple \texttt{diskbb} model. Due the degeneration between column density and iron abundance, the parameters can not be constrained well if all are setting free. Therefore, we set iron abundance to the best-fitting result of observation 4. In $\chi^2$ plots, the blue dots represent results from model without \texttt{xstar}, the orange dots are the results when \texttt{xstar} is added.}
    \label{figure11}
\end{figure*}

We then try to track the evolution of disk wind by fitting all observations. The results of well constrained observations are shown in Fig. \ref{figure11}. The ratio reached about $\sim 2.0$ in the early stage and declined with time later to $\sim 0.25$ (refer to Fig. \ref{figure8}). The relativistic disk wind can be generated when the accretion rate of source exceeds its Eddington limit. The column density and ionization log$(\xi)$ remain almost constant (within uncertainties) during observations. The ionization log$(\xi)$ remains about 5 and the column density is about $2\times10^{23}$ cm$^{-2}$. We notice that the column density is almost constant with decreasing of luminosity of source. The mass loss rate of the disk is proportional to the disk luminosity, $\dot{m}\propto L_{d}$. The dynamical timescale of wind is $t_{w}\propto v_{w}^{-1}$. The column density of absorptive wind should be $N_{H}\propto L_{d}/v_{w}$. The simulation result shows that wind velocity is not linear with disk luminosity, and a brighter disk will generate faster winds. For low disk luminosity, the winds may have acceleration-deceleration-re-acceleration process \citep{2019PASJ...71...70T}. This process will make the dynamical time scale of wind longer. Therefore, this variation of the observed column density is possible. A more detailed calculation is needed to provide a convincing explanation. The effect of magnetic field is ignored in \citet{2019PASJ...71...70T}. However, the disk wind may originate from the magnetic field around the source. GR-MHD simulations show that outflow mass flux will increase with time in some periods while the inflow mass flux decreasing. And the mass fluxes of accretion and radial outflow along the disk are both dominated by the vertical mass loss from the disk surface \citep{2019ApJ...882....2V}, which could explain the increasing or constant of $N_H$ when accretion rate is decreasing. The redshift parameter $z$ grows about 0.05 during observation period (i.e., from $-0.24$ to $-0.19$), which indicating disk wind is getting slower. Numerical simulations show that disk winds are gradually accelerated to its terminal velocity, where a higher disk luminosity will lead to a faster terminal wind velocity \citep{2001PASJ...53..285H,2019PASJ...71...70T,2019ApJ...882....2V,2021MNRAS.502.5797Y}. The source luminosity is decreasing with time which can explain the observed evolution of wind velocity.


\section{Conclusion}

In this work, we determined the moderate spin of the black hole in 4U 1543-47 using the data observed by \textit{Insight}-HXMT during the 2021 outburst. We used a simple phenomenological model and a continuum-fitting model to study the spectra in the soft state. The spin of 4U 1543-47 is $a_*=0.46\pm0.12(1\sigma)$ considering the uncertainties of the adopted parameters, e.g., the BH mass, distance, and inclination angle of the disk.

Besides, we noticed a wide absorption line at $\sim 8-10$ keV in the spectra and try to fit it with \texttt{xstar} model. A simple Gaussian model \texttt{gabs} suggested the broad feature would come from the turbulent velocity of $\sim 30000$ km s$^{-1}$. When applying \texttt{xstar} to the spectral data, we set the ionization of the plasma log$(\xi)$, the column density $N_H$, the iron abundance Z$_{\textrm Fe}$ and the redshift z as free parameters, finding that the fast wind have a velocity of $v\sim 0.2 c$ with the iron rich abundance $\textrm{Z}_{\textrm Fe}\sim 3$. We also study the evolution of disk winds from the early stage of the outburst to the end (see plots in Fig. \ref{figure11}). The column density and ionization log$(\xi)$ remain almost constant (within uncertainties) during observations. The redshift parameter $z$ grows about 0.05 during the observation period, indicating that disk winds are getting slower with the declining luminosity.

\begin{table}
     \centering
     \caption{Best-fit parameters of the model \texttt{xstar} for the example spectrum based on ObsID P030402603801.}
    \renewcommand{\arraystretch}{1.5}
    \begin{tabular}{c|c|c|c}
       \hline
       Model     & Par.              & Unit         & Value  \\
       \hline
        xstar    &  $N_H$            & $\times10^{23}$ cm$^{-2}$  & $1.40^{+0.05}_{-0.02}$ \\
                 &  log$(\xi)$         &              & $4.16^{+0.11}_{-0.02}$ \\
                 &  Z$_{\textrm Fe}$ & Z$_{\sun}$   & $3.18^{+0.02}_{-0.05}$ \\
                 &  Redshift z       &              & $-0.186^{+0.002}_{-0.003}$\\
        \hline
    \end{tabular}
   
    \label{table4}
\end{table}

\begin{table}
     \centering
     \caption{Best-fit parameters of the models \texttt{xstar} for the example spectrum based on ObsID P030402603801. The iron abundance is fixed at 3.18Z$_{\sun}$, and two log$(\xi)$ are linked.}
    \renewcommand{\arraystretch}{1.5}
    \begin{tabular}{c|c|c|c}
       \hline
       Model     & Par.              & Unit         & Value  \\
       \hline
        xstar1   &  $N_H$            & $\times10^{23}$ cm$^{-2}$  & $1.43^{+0.11}_{-0.11}$ \\
                 &  log$(\xi)$         &              & $4.47^{+0.07}_{-0.08}$ \\
                 &  Redshift z       &              & $-0.197^{+0.006}_{-0.006}$\\
        xstar2   &  $N_H$            & $\times10^{23}$ cm$^{-2}$  & $0.52^{+0.07}_{-0.07}$ \\
                 &  log$(\xi)$         &              & $4.47^{+0.07}_{-0.08}$ \\
                 &  Redshift z       &              & $-0.071^{+0.011}_{-0.011}$\\
        \hline
    \end{tabular}
    \label{table5}
\end{table}

\section*{Acknowledgements}
We are grateful to the referee for the comments and thank Dr. James Steiner for the discussion. This work is supported by the National Key Research and Development Program of China (Grants No. 2021YFA0718503), the NSFC (No. 12133007). This work has made use of data from the \textit{Insight-}HXMT mission, a project funded by the China National Space Administration (CNSA) and the Chinese Academy of Sciences (CAS).
\section*{Data Availability}
Data that were used in this paper are from the Institute of High Energy
Physics Chinese Academy of Sciences(IHEP-CAS) and are publicly available for download from the \textit{Insight-}HXMT website. The kerrbb2 model used in this work is a public code available from the website https://jfsteiner.com/wordpress/?p=55 .


\bibliographystyle{mnras}
\bibliography{bib_4U_1543-47_spin} 



\appendix



\bsp	
\label{lastpage}
\end{document}